\documentclass[bibyear]{aa} % if the references are not structured
\usepackage{natbib}
\usepackage{graphicx}
\usepackage{txfonts}
\begin{document}

\title{New deuterated species in TMC-1: Detection of CH$_2$DC$_4$H with the QUIJOTE line survey \thanks{Based on observations carried out with the Yebes 40m telescope (projects 19A003, 20A014, 20D023, and 21A011). The 40m radiotelescope at Yebes Observatory is operated by the Spanish Geographic Institute (IGN, Ministerio de Transportes, Movilidad y Agenda Urbana).}}

\author{
C.~Cabezas\inst{1},
R.~Fuentetaja\inst{1},
E.~Roueff\inst{2},
M.~Ag\'undez\inst{1},
B.~Tercero\inst{3,4},
N.~Marcelino\inst{3,4},
J.~R.~Pardo\inst{1},
P.~de~Vicente\inst{3}
and
J.~Cernicharo\inst{1}
}

\institute{Grupo de Astrof\'isica Molecular, Instituto de F\'isica Fundamental (IFF-CSIC), C/ Serrano 121, 28006 Madrid, Spain.
\email carlos.cabezas@csic.es; jose.cernicharo@csic.es
\and LERMA, Observatoire de Paris, PSL Research University, CNRS, Sorbonne Universit\'es, 92190 Meudon, France
\and Observatorio Astron\'omico Nacional (IGN), C/ Alfonso XII, 3, 28014, Madrid, Spain.
\and Centro de Desarrollos Tecnol\'ogicos, Observatorio de Yebes (IGN), 19141 Yebes, Guadalajara, Spain.
}

\date{Received; accepted}

\abstract{We report the first detection in space of the single deuterated isotopologue of methyldiacetylene, CH$_2$DC$_4$H. A total of 12 rotational transitions, with $J$ = 8-12 and $K_a$ = 0 and 1, were identified for this species in TMC-1 in the 31.0-50.4 GHz range using the Yebes 40m radio telescope. The observed frequencies allowed us to obtain, for the first time, the spectroscopic parameters of this deuterated isotopologue. We derived a column density of (5.5$\pm$0.2)$\times$10$^{11}$ cm$^{-2}$. The abundance ratio
between CH$_3$C$_4$H and CH$_2$DC$_4$H is 24$\pm$2. This ratio is similar to that found for the CH$_3$C$_3$N/CH$_2$DC$_3$N analogue system, which is 22$\pm$2. We did not detect the deuterated species CH$_3$C$_4$D, which has already been observed in laboratory experiments. The detection of deuterated CH$_3$C$_4$H allows us to extend the discussion on the chemical mechanisms of deuterium fractionation at work in TMC-1 using a new gas-phase chemical model with multiply deuterated molecules. Introducing a possible deuterium exchange reaction between CH$_3$CCH and atomic deuterium allows us to account for the CH$_3$C$_4$H/CH$_2$DC$_4$H abundance ratio.}


\titlerunning{CH$_2$DC$_4$N in TMC-1}
\authorrunning{Cabezas et al.}

\maketitle

\section{Introduction}

Studies of deuterium fractionation provide insight into the chemical reaction mechanisms that occur both in the gas phase and on the surfaces of dust particles. Deuterium fractionation also may provide us with sensitive probes for the early stages of protostellar evolution. It is therefore important not only to discover the degree of molecular deuteration possible in specific interstellar species, but also to determine this as a function of the environment. Deuterium fractionation is understood to be a very efficient process responsible for the extremely high D/H ratios found for interstellar molecules such as HDCS \citep{Marcelino2005} and CH$_2$DOH \citep{Parise2006}. Moreover, it allows multiple deuteration for abundant molecules containing more than one hydrogen atom (\citealt{Agundez2021a} and references therein).

Very sensitive broadband line surveys of astronomical sources have helped to increase the number of new molecular identifications in the last years, because weak lines arising from low-abundance species and from low-dipole moment species can now be easily detected \citep{Agundez2021b,Cernicharo2021a,Cabezas2021a,Cabezas2021b}. The downside of this high sensitivity is the large number of new
lines that appear in the survey. Deuterated isotopologues of abundant interstellar molecules significantly contribute to these new spectral features. Considering all the information mentioned above, the astronomical identification of these isotopologues is of the utmost importance, not only to gain knowledge on their molecular formation pathways or on how deuterium fractionation works, but also to assign unidentified features in line surveys.

Several single deuterated isotopologues of species such as H$_3$CN, CH$_3$CCH, $c$-C$_3$H$_2$, C$_4$H, H$_2$C$_4$, H$_2$CCN,
HC$_3$N, and HC$_5$N \citep{Cabezas2021c} have been detected in TMC-1 using the on going QUIJOTE\footnote{\textbf{Q}-band \textbf{U}ltrasensitive \textbf{I}nspection \textbf{J}ourney to the \textbf{O}bscure \textbf{T}MC-1 \textbf{E}nvironment} line survey \citep{Cernicharo2021b}. As part of the QUIJOTE line survey, we also identified the single deuterated isotopologues HDCCN \citep{Cabezas2021c} and CH$_2$DC$_3$N \citep{Cabezas2021d} in TMC-1. The identification of HDCCN in TMC-1 is supported by the observation of the rotational spectrum of that species, while that for CH$_2$DC$_3$N is based on quantum chemical calculations.

In our previous work on the CH$_2$DC$_3$N observation \citep{Cabezas2021d}, we searched for the spectral signatures of the deuterated species of methyldiacetylene (CH$_3$C$_4$H). This molecule was discovered in TMC-1 by \citet{Walmsley1984} and it has been found to be 7.5 times more abundant than CH$_3$C$_3$N \citep{Marcelino2021,Cernicharo2021c}. We followed the same strategy used for CH$_2$DC$_3$N to predict the transition frequencies of CH$_2$DC$_4$H. We found only two lines at the predicted transition frequencies, but no other lines predicted in the frequency range of the line survey were observed. We derived a 3$\sigma$ upper limit to its column density of 3.7$\times$10$^{11}$ cm$^{-2}$. Here we report the identification of spectral lines of the deuterated species CH$_2$DC$_4$H in TMC-1, based on our previous ab initio calculations and on a wider spectral search. The observed deuterium fractions are compared to an extended chemical model including the related deuterated compounds.

\section{Observations}

The data presented in this work are part of the QUIJOTE spectral line survey in the Q band towards TMC-1 ($\alpha_{J2000}=4^{\rm h} 41^{\rm  m} 41.9^{\rm s}$ and $\delta_{J2000}=+25^\circ 41' 27.0''$), which was performed at the Yebes 40m radio telescope during various observing sessions between November 2019 and April 2021. A detailed description of the QUIJOTE line survey is provided in \citet{Cernicharo2021b}. All observations were carried out using the frequency switching technique, with a frequency throw of 10\,MHz during the two first observing runs and of 8\,MHz in the later ones. This observing mode provides a noise level $\sqrt{2}$ times lower than the unfolded data, but on the other hand it produces negative spectral features at $\pm$ 10\,MHz or $\pm$ 8\,MHz of each rotational transition. These negative features can be easily identified because of their symmetric displacement by exactly the frequency throw. The selected temperature scale is $T_A^*$. The $T_{MB}$ can be easily obtained by dividing the observed $T_A^*$ by the beam efficiency. Values of $\eta_{MB}$ have been provided by \citet{Tercero2021}. Furthermore, $T_A^*$ was calibrated using two absorbers at different temperatures and the atmospheric transmission model (ATM; \citealt{Cernicharo1985, Pardo2001}).

Different frequency coverages were observed, 31.08-49.52 GHz and 31.98-50.42 GHz, which permitted us to check that no spurious ghosts were produced in the down-conversion chain in which the signal coming from the receiver was downconverted to 1-19.5 GHz, and then split into eight bands with a coverage of 2.5 GHz, each of which were analysed by the Fast Fourier Transform spectrometers. Calibration uncertainties were adopted to be 10~\% based on the observed repeatability of the line intensities between different observing runs. All data were analysed using the GILDAS package\footnote{\texttt{http://www.iram.fr/IRAMFR/GILDAS}}.

\section{Results}

\begin{table}
\small
\caption{Observationally derived and theoretical spectroscopic parameters (in MHz) for CH$_2$DC$_4$H.}
\label{rot_const}
\centering
\begin{tabular}{{lcc}}
\hline
\hline
Constant          &  Space$^a$             & Ab initio$^b$   \\
\hline
\hline
$A$               &  120899.40$^c$         &     120899.40    \\
$B$               &  1965.90898(101)       &      1965.96     \\
$C$               &  1958.01640(129)       &      1958.41     \\
$\Delta_{J}$      &  1.001(42)10$^{-4}$    &       -          \\
$\Delta_{JK}$     &  1.175(46)10$^{-2}$    &       -          \\
$rms$$^d$         &   12.2                 &       -          \\
$J_{min}/J_{max}$ &   8/12                 &       -          \\
$K_{min}/K_{max}$ &   0/1                  &       -          \\
N$^e$             &   12                   &       -          \\
\hline
\hline
\end{tabular}
\tablefoot{
        \tablefoottext{a}{Fit to the lines of CH$_2$DC$_4$H observed in TMC-1.} \tablefoottext{b}{CCSD/cc-pVTZ level of theory. Scaled values using CH$_3$C$_4$H as a reference . See text.} \tablefoottext{c}{Fixed to the calculated value.} \tablefoottext{d}{The standard deviation of the fit in kHz.} \tablefoottext{e}{Number of lines included in the fit.}
    }
\end{table}
\normalsize

\begin{table}
\tiny
\caption{Observed line parameters for CH$_2$DC$_4$H in TMC-1.}
\label{freq_lines}
\centering
\begin{tabular}{lcccc}
\hline
\hline
{$(J_{K_{\rm a},K_{\rm c}})_{\rm u}$-$(J_{K_{\rm a},K_{\rm c}})_{\rm l}$} & $\nu_{obs}$~$^a$ & $\int T_A^* dv$~$^b$ & $\Delta v$~$^c$ & $T_A^*$ \\
                     &  (MHz)              & (mK\,km\,s$^{-1}$)      & (km\,s$^{-1}$)  & (mK) \\
\hline
\hline
$ 8_{0, 8}- 7_{0, 7}$  &  31391.195   & 1.29$\pm$0.14 & 0.69$\pm$0.08 & 1.75$\pm$0.15 \\
$ 8_{1, 7}- 7_{1, 6}$  &  31422.580   & 1.10$\pm$0.23 & 0.87$\pm$0.20 & 1.20$\pm$0.13 \\
$ 9_{1, 9}- 8_{1, 8}$  &  35279.316   & 1.37$\pm$0.26 & 1.03$\pm$0.26 & 1.25$\pm$0.12 \\
$ 9_{0, 9}- 8_{0, 8}$  &  35315.051   & 1.82$\pm$0.16 & 0.78$\pm$0.08 & 2.18$\pm$0.13 \\
$ 9_{1, 8}- 8_{1, 7}$  &  35350.328   & 0.59$\pm$0.10 & 0.64$\pm$0.18 & 0.86$\pm$0.12 \\
$10_{1,10}- 9_{1, 9}$  &  39199.138   & 0.82$\pm$0.17 & 0.56$\pm$0.11 & 1.37$\pm$0.15 \\
$10_{1, 9}- 9_{1, 8}$  &  39278.077   & 0.39$\pm$0.11 & 0.52$\pm$0.14 & 0.71$\pm$0.11 \\
$11_{1,11}-10_{1,10}$  &  43118.976   & 0.59$\pm$0.08 & 0.39$\pm$0.09 & 1.39$\pm$0.13 \\
$11_{0,11}-10_{0,10}$  &  43162.660   & 1.24$\pm$0.10 & 0.48$\pm$0.05 & 2.40$\pm$0.16 \\
$11_{1,10}-10_{1, 9}$  &  43205.812   & 1.24$\pm$0.17 & 0.93$\pm$0.15 & 1.25$\pm$0.15 \\
$12_{1,12}-11_{1,11}$  &  47038.786   & 0.63$\pm$0.12 & 0.47$\pm$0.10 & 1.27$\pm$0.19 \\
$12_{0,12}-11_{0,11}$  &  47086.391   & 1.12$\pm$0.20 & 0.80$\pm$0.18 & 1.31$\pm$0.20 \\
\hline
\hline
\end{tabular}
\tablefoot{
\tablefoottext{a}{Observed frequencies towards TMC-1 for which we adopted a v$_{\rm LSR}$ of 5.83 km s$^{-1}$ \citep{Cernicharo2020a}. The
frequency uncertainty 10 kHz.}\tablefoottext{b}{Integrated line intensity in mK\,km\,s$^{-1}$.} \tablefoottext{c}{Line width at half intensity derived by fitting a Gaussian function to the observed line profile (in km\,s$^{-1}$).}
}\\
\end{table}

\begin{figure*}
\centering
\includegraphics[angle=0,width=\textwidth]{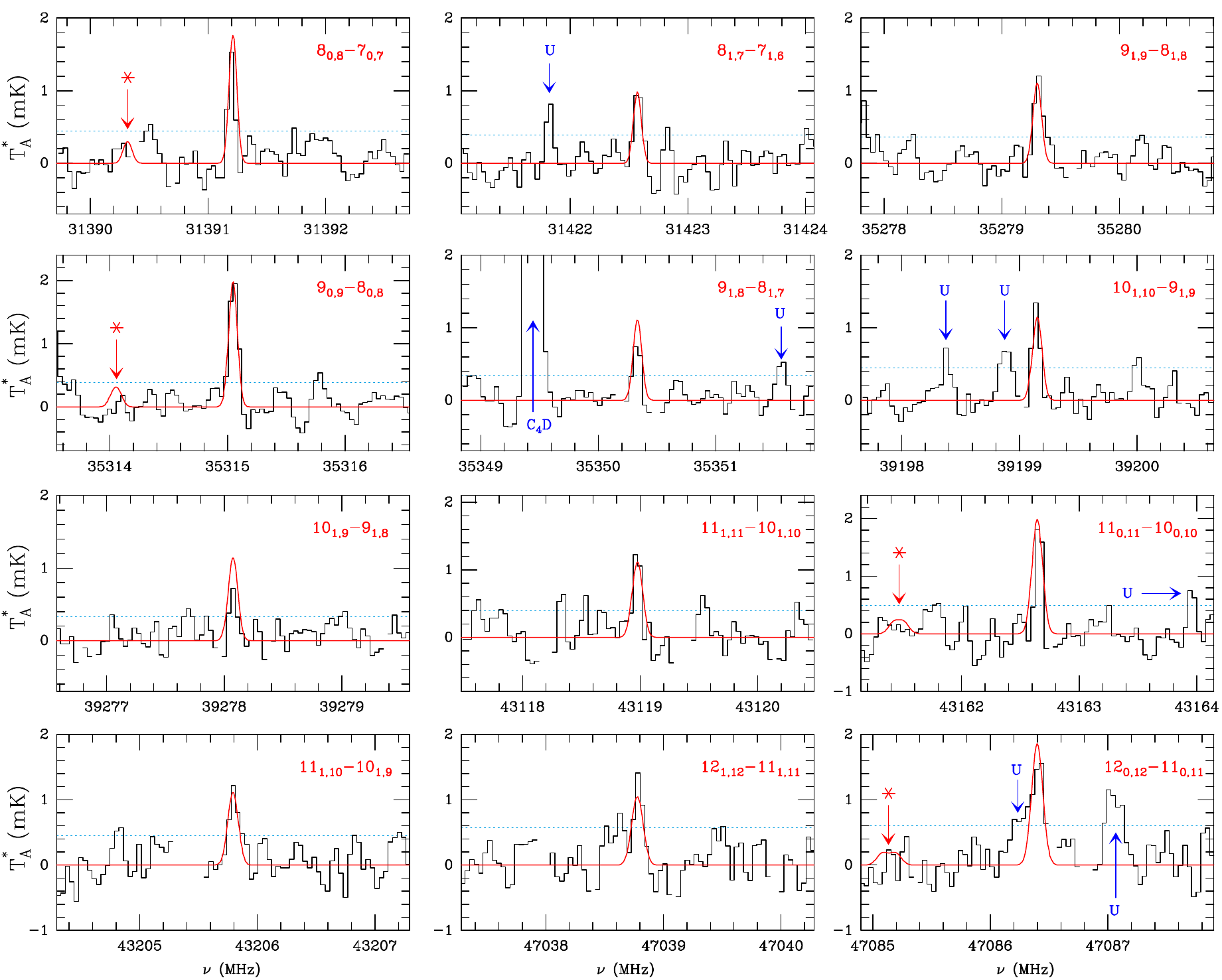}
\caption{Observed lines of CH$_2$DC$_4$H in TMC-1 in the 31.0-50.4 GHz range.
Frequencies and line parameters are given in Table \ref{freq_lines}. Quantum numbers
for the observed transitions are indicated in each panel. The red line shows the synthetic spectrum computed for a rotational temperature of 8\,K and a column density of (5.5$\pm$0.2)$\times$10$^{11}$ cm$^{-2}$ (see text); also, the dashed blue line indicates the 3$\sigma$ noise level. Blank channels correspond to negative features created during the folding of the frequency switching data. The asterisk indicates $K_a$=2 transitions and the label U corresponds to features above 4$\sigma$.} \label{spectra}
\end{figure*}

In a previous work, we reported the identification of the CH$_2$DC$_3$N species \citep{Cabezas2021d}. Among the forest of the unidentified lines in our QUIJOTE line survey, we found a series of five lines with a harmonic relation between them and two additional series of lines at higher and lower frequencies from those lines. After discarding other candidates, we assigned those lines to CH$_2$DC$_3$N. The assignment was based on the change in the rotational parameters of CH$_3$C$_3$N produced by the H–D exchange, which were obtained by ab initio calculations. As mentioned above, we already considered that the rotational transitions of CH$_2$DC$_4$H should also be observed in our line survey. As for the CH$_3$C$_3$N/CH$_2$DC$_3$N system, we performed geometry optimisation calculations for CH$_3$C$_4$H and CH$_2$DC$_4$H in order to estimate the isotopic shift on the rotational constants for the  CH$_3$C$_4$H/CH$_2$DC$_4$H species. We employed the CCSD/cc-pVTZ level of theory \citep{Cizek1969,Dunning1989} which reproduces the $B$ rotational constant well for CH$_3$C$_4$H, 2027.4 MHz versus the experimental value of 2035.74 MHz. The theoretical values for the rotational constants of CH$_2$DC$_4$H were then scaled using the experimental/theoretical ratio obtained for CH$_3$C$_4$H, and the results are shown in Table \ref{rot_const}. Using these rotational constants, we predicted the transition frequencies for CH$_2$DC$_4$H and we observed only two lines at those frequencies with intensities of $\sim$ 1\,mK in $T_A^*$. We searched for other lines predicted in the frequency range of the line survey, but we could not observed them. We assumed that the lines for CH$_2$DC$_4$H species could not be detected with the sensitivity achieved at that moment.

A deeper inspection of the frequency regions where the CH$_2$DC$_4$H lines are predicted reveals the presence of a series of lines that we did not consider in our first analysis. In our previous search, we focussed on two lines with intensities of $\sim$ 1 mK at 31382.4 and 35304.7 MHz that agree well with the predicted frequencies. We incorrectly assigned them to the 8$_{0,8}$-7$_{0,7}$ and 9$_{0,9}$-8$_{0,8}$ transitions of CH$_2$DC$_4$H since the deviation from predictions were similar to that found for the lines of CH$_2$DC$_3$N. The new detected lines corresponding to these transitions appear at higher frequencies, 31391.2 and 35315.0 MHz. Using these two lines as a guide, we also observed another two lines for the $J$+1$_{0,J+1}$$\leftarrow$$J_{0,J}$ series and eight transitions for the $J$+1$_{1,J+1}$$\leftarrow$$J_{1,J}$ and  $J$+1$_{1,J}$$\leftarrow$$J_{1,J-1}$ progressions. The rotational transitions 10$_{0,10}$-9$_{0,9}$ and 12$_{1,11}$-11$_{1,10}$ were not included in the fit because they were detected below the 3$\sigma$ noise level. On the other hand, $K_a$=2 transitions are too weak to be detected with the current sensitivity (see Fig. \ref{spectra}). All the observed lines, shown in Table \ref{freq_lines} and Fig. \ref{spectra}, were analysed using an asymmetric rotor Hamiltonian in the FITWAT code \citep{Cernicharo2018} to derive the rotational and centrifugal distortion constants shown in Table \ref{rot_const}. As it can be seen, the predicted values for CH$_2$DC$_4$H perfectly agree with those derived from our fit, which allows us to conclude that the spectral carrier of our lines is CH$_2$DC$_4$H. It should be noted that the calculations provide the equilibrium values for the rotational constants ($A_e$, $B_e$, and $C_e$), while the experimental values are the ground state rotational constants ($A_0$, $B_0$, and $C_0$). Despite the equilibrium rotational constants slightly differing from the ground state constants, we can assume similar discrepancies for CH$_3$C$_4$H and CH$_2$DC$_4$H and, thus, the estimated constants for CH$_2$DC$_4$H are essentially unaffected by this fact. Using experimental/theoretical ratios is the most common method to predict the expected experimental rotational constants for an isotopic species of a given molecule when the rotational constants for its parent species are known. The ($B+C$) calculated value shows a relative error from the experimental value of 0.01\%, similar to what was found in the case of CH$_2$DC$_3$N. With the available data for CH$_2$DC$_4$H, we cannot determine the experimental value for the $A$ rotational constant, which was kept fixed to the ab initio value, as it was done for CH$_2$DC$_3$N.

The synthetic spectrum of CH$_2$DC$_4$H was computed assuming a dipole moment value identical to that of the main isotopologue  CH$_3$C$_4$H (1.207\,D; \citealt{Bester1984}) and the partition function shown in Table \ref{pfunction}. The column density of CH$_2$DC$_4$H was derived from a rotational diagram analysis of the observed intensities. We considered a source of uniform brightness with a radius of 40$''$ \citep{Fosse2001}. We derived $T_r$=8.0$\pm$0.5\,K and  N(CH$_2$DC$_4$H)=(5.5$\pm$0.2)$\times$10$^{11}$ cm$^{-2}$. This value is not far from the 3$\sigma$ upper limit to the CH$_2$DC$_4$H column density of 3.7$\times$10$^{11}$ cm$^{-2}$ that we provided in our previous work  \citep{Cabezas2021d}. As shown in Fig. \ref{spectra}, the agreement between the synthetic spectrum and the observations is excellent. The column density is not very sensitive to the adopted value of the rotational temperature between 6 and 10\,K. The column density for the parent isotopologue CH$_3$C$_4$H was derived by \citet{Cernicharo2021c} to be (1.30$\pm$0.04)$\times$10$^{13}$ cm$^{-2}$. Therefore, the CH$_3$C$_4$H/CH$_2$DC$_4$H abundance ratio is 24$\pm$2. This ratio is similar to that of 22$\pm$2 found for the CH$_3$C$_3$N/CH$_2$DC$_3$N analogue system. For the deuterated species CH$_3$C$_4$D, for which laboratory spectroscopy is available \citep{Heath1955}, we derived a 3$\sigma$ upper limit to its column density of 9$\times$10$^{10}$ cm$^{-2}$. Hence,
N(CH$_3$C$_4$H)/N(CH$_3$C$_4$D)$\ge$144.

\begin{table}
\begin{center}
\caption[]{Rotational partition function for CH$_2$DC$_4$H at different temperatures calculated at a maximum value of $J$=70.}
\scalebox{1}{
\label{pfunction}
\begin{tabular}{cc}
\hline
\hline
Temperature/K   &\hfill $Q_r$\hfill\mbox{}\\
\hline
   9.375          &     225.0    \\
   18.750         &     635.5    \\
   37.500         &     1796.1   \\
   75.000         &     5069.2   \\
   150.000      &       13704.1  \\
   225.000      &       22769.6  \\
   300.000      &       30972.2  \\
\hline
\end{tabular}
}
\end{center}
\end{table}

\section{Chemical modelling}

\begin{table*}
\small
\caption{Deuteration fractions in TMC-1   compared to our gas phase chemical model.}
\label{tab_H/D}
\centering
\begin{tabular}{{lccccc}}
\hline
\hline
Molecules  &  TMC-1 &     Model A$^a$   & Model A$_n$ $^b$ &  Model B$^a$& Model B$_n$ $^b$ \\
\hline
\hline
CH$_3$C$_4$H/CH$_2$DC$_4$H  &        $24^b$   & 59      &       31      &       20      &       16  \\
CH$_3$C$_4$H/CH$_3$C$_4$D   &  $\ge$144$^c$   & 134     &       78      &       55      &       44  \\
CH$_3$C$_3$N/CH$_2$DC$_3$N  &      $ 22^a $   & 76      &       29      &       31      &       22  \\
CH$_3$CN/CH$_2$DCN          &       11$ ^d$   & 15      &       15      &       15      &       15  \\
H$_2$CCN/HDCCN              &        20$^d$   & 24      &       24      &       24      &       24  \\
HC$_3$N/DC$_3$N             &        62$^e$   & 55      &       50      &       55      &       52  \\
HNCCC/DNCCC                 &        43$^e$   & 35      &       34      &       35      &       34  \\
HCCNC/DCCNC                 &        30$^e$   & 27      &       27      &       27      &       27  \\
HC$_5$N/DC$_5$N$^b$         &        82$^e$   & 23      &       20      &       23      &       20  \\
$c$-C$_3$H$_2$/$c$-C$_3$HD  &        27$^d$   & 46      &       36      &       45      &       36  \\
C$_4$H/C$_4$D               &       118$^d$   & 55      &       46      &       55      &       46  \\
H$_2$C$_4$/HDC$_4$          &        83$^d$   & 50      &       33      &       49      &       33  \\
CH$_3$CCH/CH$_3$CCD         &        49$^d$   & 257     &       46      &       264     &       48  \\
CH$_3$CCH/CH$_2$DCCH        &        10$^d$   & 76      &       15      &       76      &       15  \\
\hline
\hline
\end{tabular}
\tablefoot{ \tablefoottext{a}{\citet{Cabezas2021d}.} \tablefoottext{b}{This work.} \tablefoottext{c}{3$\sigma$ upper limit.} \tablefoottext{d}{\citet{Cabezas2021c}.} \tablefoottext{e}{\citet{Cernicharo2020a}.}\\
  }
\end{table*}
\normalsize

This new detection of deuterated CH$_3$C$_4$H allows us to extend the discussion on the chemical mechanisms of deuterium fractionation at work in TMC-1. We would like to point out that our results have been obtained with the same assumed physical conditions for the TMC-1 environment as in our previous studies \citep{Cabezas2021c,Cabezas2021d}. In Table \ref{tab_H/D}, we display the different deuterium fractions found in TMC-1 so far, as well as the previous steady state results corresponding to models A and B of \cite{Cabezas2021d}. We recall that a full scrambling scenario of the reactions between C$_2$D and CH$_3$CCH (model B) had predicted a CH$_3$C$_4$H/CH$_2$DC$_4$H ratio of 20, which is close to the value derived with the present detection. In model A, these reactions were assumed to proceed directly, that is
\begin{equation}
\rm{C_2D + CH_3CCH \rightarrow CH_3C_4D + H}
\end{equation}
and
\begin{equation}
\rm{C_2H + CH_2DCCH \rightarrow CH_2DC_4H + H},
\end{equation}
resulting in a much larger CH$_3$C$_4$H/CH$_2$DC4H ratio.
The relevant chemical processes principally involve the reactions of  C$_2$D with methylacetylene CH$_3$CCH and the allene isomer CH$_2$CCH$_2$ as well as the reactions between C$_2$H with their deuterated substitutes CH$_3$CCD, CH$_2$DCCH,  and CHDCCH$_2$. Dissociative recombination reactions of CH$_2$DC$_4$H$_2^+$ and CH$_3$C$_4$HD$^+$ may also contribute to deuterated CH$_3$C$_4$H as ion-molecule reactions between H$_2$D$^+$ and CH$_3$C$_4$H are leading to these complex molecular ions. This is in contrast with  CH$_3$CCH  or CH$_2$CCH$_2$ and H$_2$D$^+$ reactions,  which result in the ejection of two hydrogen molecules, and the formation of the linear and cyclic C$_3$H$_3^+$ ions as reported in the experiments of \cite{abeysekera:15}.

We would like to point out that deuterated methylacetyne is underproduced compared to the hydrogenated compound in all chemical scenarios considered so far. We thus introduced a possible deuterium exchange reaction between CH$_3$CCH and atomic deuterium:
\begin{equation}
{\rm{D} + CH_3CCH \rightarrow CH_2DCCH + H}, ~~k_1=2 \times 10^{-11} cm^3s^{-1}
\end{equation}
and
\begin{equation}
{\rm{D} + CH_3CCH \rightarrow CH_3CCD + H}, ~~k_2=0.66 \times 10^{-11} cm^3s^{-1}
.\end{equation}

We used a modest value for the corresponding reaction rate coefficients consistent with values expected for neutral-neutral reactions without an activation barrier and assumed that the channel to CH$_3$CCD is one-third of that of CH$_2$DCCH, following statistical arguments. The computed deuterium fractions are reported in column A$_n$  and B$_n$  of Table \ref{tab_H/D}. This single change, introduced in the previous reported models, allowed us to obtain good agreement for the deuterium fractions of both isotopologues of methyl acetylene. The ratio for methylcyanoacetylene, CH$_3$C$_3$N/CH$_2$DC$_3$N, was reduced as well, allowing for a satisfactory  comparison with observations. We see that the values of the CH$_3$C$_4$H/CH$_2$DC$_4$H ratio are also consequently somewhat reduced, as expected. The values corresponding to A  and B models bracket the observational values now.  We shall also point out that both models predict a smaller than observed ratio for CH$_3$C$_4$H/CH$_3$C$_4$D. We feel that the level of agreement between observations and models A$_n$ and B$_n$ is nevertheless quite supportive of a reasonable understanding of the deuterium enhancement in the TMC-1 environment. Further experimental and theoretical studies of the CH$_3$CCH exchange reaction with atomic deuterium would be very useful in order to probe the relevance of our hypothesis.

We acknowledge the remaining uncertainties on the chemical network. We did not introduce allenyl acetylene, H$_2$CCCHCCH, a new C$_5$H$_4$ isomer detected recently by \cite{Cernicharo2021c} in TMC-1, but we focussed our efforts on understanding the possible deuterium enhancement scenarios. This particular attempt did not allow to discriminate if the reactions between C$_2$H (C$_2$D) with CH$_3$CCH (CH$_3$CCD, CH$_2$DCCH) proceed directly or involve the formation of  a temporary intermediate complex redistributing the different stable deuterated compounds. The significant deuterium enhancement in methylacetylene obtained with the assumed deuterium exchange reaction allowed us to account for CH$_3$C$_4$H/CH$_2$DC$_4$H. We also found that ion-molecule reactions involving H$_2$D$^+$, which were introduced in the present chemical model, do not play a significant role  for both the deuteration of CH$_3$CCH and CH$_3$C$_4$H, as reactions with C$_2$H (C$_2$D) are more efficient. An additional test of our assumptions would be the observational derivation of the C$_2$H/C$_2$D ratio which is predicted to be about 7 in our models. Transition frequencies of CCD have been recently updated by \cite{Cabezas2021e}, and CCD has been detected in TMC-1 using IRAM 30m data. A derivation and discussion of the CCH/CCD ratio in TMC-1 is currently in process and will be presented in the near future in a separate article.

\section{Conclusions}

We have presented the first detection in space of a new single deuterated compound derived from methyldiacetylene, CH$_2$DC$_4$H, towards the dark cloud TMC-1. Using the Yebes 40m radio telescope, we observed a total of 12 rotational transitions, with $J$ = 8-12 and $K_a$ = 0 and 1, in the 31.0-50.4 GHz range. These transitions were assigned to CH$_2$DC$_4$H based on the excellent agreement found between the ab initio molecular constants and those derived from the frequency transitions fit. We derived a column density of (5.5$\pm$0.2)$\times$10$^{11}$ cm$^{-2}$ and a CH$_3$C$_4$H/CH$_2$DC$_4$H abundance ratio of 24$\pm$2. The ratio is similar to that of 22$\pm$2 found for the CH$_3$C$_3$N/CH$_2$DC$_3$N analogue system. We have constructed a new gas-phase chemical model, including
multiply deuterated molecules and introducing a possible deuterium exchange reaction between CH$_3$CCH and atomic deuterium. This new model is able to reproduce the abundance ratio for CH$_3$C$_4$H/CH$_2$DC$_4$H and also that for CH$_3$C$_3$N/CH$_2$DC$_3$N.

\begin{acknowledgements}
We thank ERC for funding through grant ERC-2013-Syg-610256-NANOCOSMOS. The Spanish authors thank Ministerio de Ciencia e Innovaci\'on for funding support through projects PID2019-106235GB-I00 and PID2019-107115GB-C21 / AEI / 10.13039/501100011033. MA thanks Ministerio de Ciencia e Innovaci\'on for grant RyC-2014-16277. ER acknowledges the support of the Programme National 'Physique et Chimie du Milieu Interstellaire' (PCMI) of CNRS/INSU with INC/INP co-funded by CEA and CNES.
Several kinetic data we used have been taken from the online databases KIDA (\cite{Wakelam2012}, http://kida.obs.u-bordeaux1.fr)
and UMIST2012 (\cite{McElroy2013}, http://udfa.ajmarkwick.net).
\end{acknowledgements}

\end{document}